\documentclass[a4paper,fleqn,usenatbib]{mnras}

\usepackage{mathptmx}

\usepackage[T1]{fontenc}
\usepackage{ae,aecompl}

\usepackage{graphicx,epstopdf}	
\usepackage{amsmath}	
\usepackage{amssymb}	

\defcitealias{Ginzburg2016}{GSS16}

\title[Tidal Heating of Super-Earths]{Tidal Heating of Young Super-Earth Atmospheres}

\author[S. Ginzburg and R. Sari]{
Sivan Ginzburg\thanks{E-mail: sivan.ginzburg@mail.huji.ac.il}
and Re'em Sari
\\
Racah Institute of Physics, The Hebrew University, Jerusalem 91904, Israel}

\date{Accepted XXX. Received YYY; in original form ZZZ}

\pubyear{2016}

\begin{document}
\label{firstpage}
\pagerange{\pageref{firstpage}--\pageref{lastpage}}
\maketitle

\begin{abstract}
Short-period Earth to Neptune size exoplanets (super-Earths) with voluminous gas envelopes seem to be very common. These gas atmospheres are thought to have originated from the protoplanetary disk in which the planets were embedded during their first few Myr. The accretion rate of gas from the surrounding nebula is determined by the ability of the gas to cool and radiate away its gravitational energy. Here we demonstrate that heat from the tidal interaction between the star and the young (and therefore inflated) planet can inhibit the gas cooling and accretion. Quantitatively, we find that the growth of super-Earth atmospheres halts for planets with periods of about 10 days, provided that their initial eccentricities are of the order of 0.2. Thus, tidal heating provides a robust and simple mechanism that can simultaneously explain why these planets did not become gas giants and account for the deficit of low-density planets closer to the star, where the tides are even stronger. We suggest that tidal heating may be as important as other factors (such as the nebula's lifetime and atmosphere evaporation) in shaping the observed super-Earth population.
\end{abstract}




\section{Introduction}\label{sec:introduction}

Low density super-Earths seem to be among the most abundant planets found by the {\it Kepler} mission \citep{WolfgangLopez2015}. These are short-period Earth to Neptune sized planets with low densities that rule out a purely rocky composition. The low density may indicate either a water-rich composition or a voluminous gas atmosphere, which is the only possibility for many extremely low-density planets \citep{Lopez2012,Lissauer2013}. 

Rocky cores can gravitationally accrete substantial hydrogen/helium atmospheres from the gas-rich protoplanetary disk that surrounds young stars. The gas accretion rate is determined by the time it takes the atmosphere to radiate away its gravitational energy (Kelvin--Helmholtz contraction). Therefore, the gas mass each rocky core acquires from the surrounding nebula is determined by comparing the atmosphere's cooling time to the nebula's lifetime \citep[a few Myr; see, e.g.,][]{Mamajek2009,WilliamsCieza2011,Alexander2014}. If the core obtains an atmosphere that is comparable to its own mass, the accretion rate of additional gas increases and the planet quickly evolves into a gas giant by runaway growth \citep{BodenheimerPollack86,Pollack96,PisoYoudin2014,Piso2015}.

Recent studies have found that the gas accretion rate onto short-period super-Earths is sufficient to obtain atmospheres of a few percent in mass, in accordance with their observed low densities \citep{Lee2014,InamdarSchlichting2015,LeeChiang2015,Ginzburg2016}. \citet{Lee2014} even suggest that the accretion rate is too high, transforming the planets into gas giants (Jupiters) instead of super-Earths. \citet{InamdarSchlichting2015} raise the opposite concern and argue that the formation of super-Earth cores by collisions of smaller protoplanets removes gas atmospheres and delays further gas accretion until the gas disk begins to disperse, resulting in lighter atmospheres. \citet{InamdarSchlichting2015} and \citet{LeeChiang2016} show that these short-lived and gas-depleted environments give rise to super Earths with a few percent by mass gas envelopes. 

The atmosphere mass when the nebula disperses is not necessarily the mass that we infer today. Atmospheric mass can be lost following the nebula's dispersal due to stellar UV irradiation \citep{Rogers2011,Lopez2012,OwenJackson2012,LopezFortney2013,OwenWu2013} or spontaneously, due to the loss of pressure support from the surrounding disk combined with heating from the inner layers of the atmosphere or from the rocky core \citep{IkomaHori2012,Ginzburg2016,OwenWu2016}. These mechanisms, especially evaporation due to UV irradiation, are often invoked to explain the scarcity of low density super-Earths in very close orbits. 

Here, we discuss a different mechanism that can prevent the accretion of heavy gas atmospheres in the first place. As explained above, atmosphere accretion is equivalent to (Kelvin--Helmholtz) cooling. Therefore, a constant supply of heat can impede the atmosphere's cooling, resulting in lighter atmospheres. One possible heat source is the heat generated by planetesimal impacts \citep{Pollack96,Rafikov2006,Rafikov2011}. However, this is an unlikely source in the inner disk because planetesimals are consumed on a short timescale \citep{GLS} and because the solid accretion rate required to stop the atmosphere's cooling might be too high \citep{LeeChiang2015}. We suggest tidal heating as an alternative heat source. Dissipation of tidal energy has been proposed as a mechanism to inhibit the cooling of hot Jupiters, thus explaining their anomalously inflated radii \citep{Bodenheimer2001,Bodenheimer2003,Gu2003,WinnHolman2005,Jackson2008,Liu2008,IbguiBurrows2009,Miller2009,Ibgui2010,Ibgui2011,Leconte2010}. In the context of super-Earths, tidal dissipation has the potential to limit the atmosphere's cooling (accretion), thus preventing runaway growth into gas giants and explaining the deficit of low density super-Earths at close separations from the star, regardless of stellar irradiation.

The outline of the paper is as follows. In Section \ref{sec:structure} we present our model for planets during the nebular gas accretion phase. In Section \ref{sec:heating} we calculate the heat dissipated by tides and in Section \ref{sec:effect} we discuss how this heat affects gas accretion. Section \ref{sec:observations} relates our theory to the observed super-Earth population and our conclusions are summarized in Section \ref{sec:conclusions}.

\section{Accreting Planet Structure}\label{sec:structure}

In this section we briefly describe the structure of the planet and its atmosphere during gas accretion from the surrounding nebula. A more detailed description is given in \citet{Ginzburg2016}, hereafter \citetalias{Ginzburg2016}, based on previous analytical and numerical studies \citep{Rafikov2006,Lee2014,PisoYoudin2014,LeeChiang2015,Piso2015}.

We assume a rocky core of mass $M_c$ and radius $R_c$ with a gas envelope of mass $M_{\rm atm}\ll M_c$ embedded inside a gas disk with an ambient density $\rho_d$ and temperature $T_d$. The atmosphere of the planet is assumed to blend into the surrounding nebula at a radius $R_{\rm out}=\min(R_{\rm B},R_{\rm H})\gg R_c$, with $R_{\rm B}$ and $R_{\rm H}$ denoting the Bondi and Hill radii, respectively. As it cools, the atmosphere develops a radiative envelope that is connected to the convective interior at the radiative--convective boundary (RCB), at a radius $R_{\rm rcb}\lesssim R_{\rm out}$. Hydrostatic equilibrium sets the following adiabatic temperature profile in the convective part of the atmosphere
\begin{equation}\label{eq:t_convective}
\frac{T(r)}{T_{\rm rcb}}=1+\frac{R_{\rm B}'}{r}-\frac{R_{\rm B}'}{R_{\rm rcb}},
\end{equation}
where $R_{\rm B}'$ is given by
\begin{equation}\label{eq:bondi}
R_{\rm B}'\equiv\frac{\gamma-1}{\gamma}\frac{GM_c\mu}{k_{\rm B}T_{\rm rcb}}\sim R_{\rm B}.
\end{equation}
Here $G$ is the gravitation constant, $k_{\rm B}$ is Boltzmann's constant, $\mu$ the molecular mass, and $\gamma$ the adiabatic index. Assuming power-law opacities, the temperature at the RCB is $T_{\rm rcb}\sim T_d$ up to an order of unity factor \citep{Rafikov2006,PisoYoudin2014,LeeChiang2015}. See however \citet{Lee2014} who take a constant $T_{\rm rcb}\approx 2500\textrm{ K}$ due to $\textrm{H}_2$ dissociation in a dusty atmosphere. The adiabatic density profile of the convective region is
\begin{equation}\label{eq:rho_convective}
\frac{\rho(r)}{\rho_{\rm rcb}}=\left(1+\frac{R_{\rm B}'}{r}-\frac{R_{\rm B}'}{R_{\rm rcb}}\right)^{1/(\gamma-1)},
\end{equation}
with $\rho_{\rm rcb}$ marking the density at the RCB.
The radius and density of the RCB are related to the outer boundary condition through the roughly isothermal profile of the radiative envelope between $R_{\rm rcb}$ and $R_{\rm out}$
\begin{equation}\label{eq:radius_rho_d}
\ln\frac{\rho_{\rm rcb}}{\rho_d}=\frac{R_{\rm B}}{R_{\rm rcb}}-\frac{R_{\rm B}}{R_{\rm out}}.
\end{equation}
Initially $\rho_{\rm rcb}=\rho_d$ and $R_{\rm rcb}=R_{\rm out}$ but as the atmosphere cools (and thereby accretes mass, as explained below) $\rho_{\rm rcb}$ increases and the radiative envelope thickens $R_{\rm rcb}<R_{\rm out}$. Nevertheless, $R_{\rm rcb}\sim R_{\rm out}$ even as the atmosphere cools ($\rho_{\rm rcb}\gg\rho_d$), due to the logarithmic dependence in Equation \eqref{eq:radius_rho_d}.

The mass of the atmosphere is given by integrating Equation \eqref{eq:rho_convective}. For $\gamma>4/3$, the mass is concentrated at the outer edge of the convective region $r\sim R_{\rm rcb}$:
\begin{equation}\label{eq:mass_atm_outside}
M_{\rm atm}\approx A(\gamma)4\pi R_{\rm rcb}^3\rho_{\rm rcb}\left(\frac{R_{\rm B}'}{R_{\rm rcb}}\right)^{1/(\gamma-1)},
\end{equation}
where the numerical coefficient is $A(\gamma)=5\pi/16$ for the diatomic $\gamma=7/5$. For $\gamma<4/3$, as \citet{LeeChiang2015} and \citet{Piso2015} effectively choose due to $\textrm{H}_2$ dissociation, the mass of the atmosphere is concentrated near its inner boundary $r\sim R_c$ and is given by
\begin{equation}\label{eq:mass_atm_inside}
M_{\rm atm}\approx\frac{\gamma-1}{4-3\gamma}4\pi R_c^3\rho_{\rm rcb}\left(\frac{R_{\rm B}'}{R_c}\right)^{1/(\gamma-1)}.
\end{equation}
We note that the radiative (roughly isothermal) portion of the atmosphere does not contribute significantly to its mass because of its exponentially declining density profile over a small scale height $h\sim R_{\rm rcb}^2/R_{\rm B}<R_{\rm rcb}$.

As the atmosphere cools, its energy $E_{\rm atm}\propto-GM_cM_{\rm atm}$ decreases, and therefore its mass and density $M_{\rm atm}\propto\rho_{\rm rcb}$ increase. Intuitively, the infalling gas radiates away its gravitational energy \citepalias[see][and references within for details]{Ginzburg2016}. We note that the rocky core does not contribute to the energy balance of the cooling atmosphere because, as seen in Equation \eqref{eq:t_convective}, its surface remains at a constant temperature of
\begin{equation}\label{eq:tmp_core}
k_{\rm B}T(R_c)=\frac{\gamma-1}{\gamma}\frac{GM_c\mu}{R_c}
\end{equation}
as long as $R_c\ll R_{\rm rcb}$ (we assume that the core is molten and fully convective at this temperature). The cooling rate (internal luminosity $L$) is determined by diffusion through the radiative layer. By combining hydrostatic equilibrium with the diffusion equation at the RCB we find
\begin{equation}\label{eq:lum}
L=\frac{64\pi}{3}\frac{\sigma T_{\rm rcb}^4R_{\rm B}'}{\kappa\rho_{\rm rcb}},
\end{equation}
with $\sigma$ denoting the Stephan--Boltzmann constant and $\kappa$ the opacity at the RCB. This result can be intuitively understood by $L\sim\sigma T_{\rm rcb}^4 R_{\rm rcb}^2/\tau$, with $\tau\sim\kappa\rho_{\rm rcb}h$ denoting the optical depth at the RCB, and $h$ the scale height there. We notice that as $M_{\rm atm}$ increases, the atmosphere becomes more optically thick and the cooling rate decreases according to Equation \eqref{eq:lum}. 

\section{Tidal Heating}\label{sec:heating}

In this section we calculate the heat dissipated by tides in the atmosphere. We first briefly review the classical problem of tidal dissipation inside a planet of roughly uniform density \citep[e.g.][]{GoldreichSoter1966}.

\subsection{Uniform Planet}\label{subsec:uniform}

The height $z$ of the tides raised on a planet with a mass $M$ and radius $R$ by a star at a distance $a$ with mass $M_\odot$ is found by equating the tidal potential $(GM_\odot/a^2)(R/a)R$ to that of the planet $GMz/R^2$: 
\begin{equation}\label{eq:tide_height}
\frac{z}{R}=\left(\frac{R}{a}\right)^3\frac{M_\odot}{M},
\end{equation}
where we have assumed Love numbers of order unity (a fluid planet). The energy stored in the tidal bulge is $GM^2z^2/R^3$, where the uniform density comes into play in the assumption that roughly $z/R$ of the planet's mass is in the tidal bulge.

We distinguish between two tidal effects: the synchronization of the planet's rotation with its orbit (tidal locking) and the circularization of eccentric orbits. As we show below, only the latter can affect super-Earth atmosphere accretion, so we leave the discussion of synchronization to Appendix \ref{sec:appendix}. In particular, we find that the synchronization timescale $t_{\rm syn}$ is short, leading us to assume a synchronized planet in an eccentric orbit with eccentricity $e\ll 1$. 
 
In each orbit, the separation of the planet from the star changes by $\Delta a/a=e$, causing a change of $\Delta z/z= e$ to the height of the tide. The corresponding energy change of the tidal bulge (relative to equilibrium) is given by $\Delta E=GM^2z^2e^2/R^3$. The dissipated power associated with the oscillation of the tidal bulge is given by
\begin{equation}\label{eq:l_circ}
L_{\rm circ}=Q^{-1}\frac{\Delta E}{P}=\frac{63\pi}{2}\frac{e^2}{QP}\frac{GM_\odot^2}{R}\left(\frac{R}{a}\right)^6,
\end{equation}
where $Q^{-1}\ll 1$ is the fraction of the oscillation energy that the damping term dissipates over a single cycle and $P$ is the orbital period. The numerical coefficient in Equation \eqref{eq:l_circ} is given by \citet{GoldreichSoter1966}.

We assume that the orbital angular momentum $M\sqrt{GM_{\odot}a(1-e^2)}$ remains constant (the planet's spin is too small to violate this assumption). We also follow previous studies \citep{Rasio1996, Gu2003} and ignore tidal changes to the spin of the star, due to their long timescale. From the conservation of angular momentum, and since the energy is given by $-GM_\odot M/(2a)$, we find the energy change due to circularization
\begin{equation}\label{eq:e_circ}
E_{\rm circ}=\frac{GM_\odot M}{2a}e^2.
\end{equation}
The circularization timescale is given by
\begin{equation}\label{eq:t_circ_uniform}
t_{\rm circ}\equiv\frac{e}{\dot{e}}=\frac{2E_{\rm circ}}{L_{\rm circ}}=\frac{2}{63\pi}Q\frac{M}{M_\odot}\left(\frac{a}{R}\right)^5P.
\end{equation}

\subsection{Extended Atmosphere}\label{subsec:extended_atmosphere}

Young super-Earths deviate significantly from the uniform density model of Section \ref{subsec:uniform}. Specifically, most of their mass $M_c$ is concentrated in a small rocky core (assumed to be of roughly uniform density), while only a small fraction $f\equiv M_{\rm atm}/M_c$ extends to large radii $R_{\rm rcb}\gg R_c$. Here and in Section \ref{subsec:young_super_earth} we calculate the tidal dissipation inside a young super-Earth, whose structure we described in Section \ref{sec:structure}, and for planets with extended atmospheres in general.

We start, for simplicity, with an equation of state with $\gamma>4/3$ for which the atmosphere mass is concentrated at $R_{\rm rcb}$. Modifications for $\gamma<4/3$ are discussed below. In this case, by repeating the derivation of Equation \eqref{eq:l_circ}, we find that the energy dissipated in the atmosphere is
\begin{equation}\label{eq:l_atm}
L_{\rm circ}=\frac{63\pi}{2}f\frac{e^2}{QP}\frac{GM_\odot^2}{R_{\rm rcb}}\left(\frac{R_{\rm rcb}}{a}\right)^6\propto M_{\rm atm}R_{\rm rcb}^5.
\end{equation}
Note that the height of the tides, given by Equation \eqref{eq:tide_height}, is inversely proportional to $M_c$ for the atmosphere as well because the core dictates the gravity (as long as $M_{\rm atm}\ll M_c$). 
For an Earth-mass ($M_\oplus$) planet in a 0.1 AU orbit around a Sun-like star $R_{\rm rcb}\approx 0.5R_{\rm B}\approx 0.5R_{\rm H}\approx 10 R_c$ \citep[see, e.g.,][and the discussion in Section \ref{subsec:young_super_earth}]{Rafikov2006,Lee2014}. Using these numbers and Equation \eqref{eq:l_atm}, we find that the ratio of the tidal power deposited in the atmosphere to that in the core is given by $f(R_{\rm rcb}/R_c)^5(Q_c/Q)$, with $Q_c\sim 10^2$ marking the rocky core's tidal quality factor \citep[based on the $Q$ value of the terrestrial planets, e.g.][but note that $Q_c$ may be larger for massive super Earths with molten cores]{GoldreichSoter1966}. We adopt a tidal dissipation factor $Q\sim 10^5$ suitable for the gas atmosphere \citep{GoldreichSoter1966,Yoder1979,BanfieldMurray1992}. 
As a consequence, tidal dissipation is dominated by the atmosphere for $f\gtrsim 1\%$ that is required to explain the observations \citep{Lopez2012}.
We therefore substitute $L_{\rm circ}$ from Equation \eqref{eq:l_atm} in Equation \eqref{eq:t_circ_uniform} and find that
\begin{equation}\label{eq:t_circ_atmosphere}
t_{\rm circ}=\frac{2}{63\pi}\frac{Q}{f}\frac{M_c}{M_\odot}\left(\frac{a}{R_{\rm rcb}}\right)^5P.
\end{equation}

\subsection{Young Super-Earth}\label{subsec:young_super_earth}

Our next step is to replace $R_{\rm rcb}\sim R_{\rm out}$ (see Section \ref{sec:structure}). What is the outer boundary $R_{\rm out}=\min(R_{\rm B},R_{\rm H})$? Coincidentally, for a few-$M_\oplus$ planet on a $\sim 0.1\textrm{ AU}$ orbit, $R_{\rm H}\sim R_{\rm B}$, so both choices yield similar results. More precisely, as long as $M_c>5M_\oplus$ and $a<0.4\textrm{ AU}$ (or equivalently $T_d\gtrsim 500\textrm{ K}$), $R_{\rm H}\lesssim R_{\rm B}$ \citep{Lee2014,InamdarSchlichting2015}. Since this range covers almost all the relevant observations \citepalias[see, e.g.,][Figure 3]{Ginzburg2016} we assume below that $R_{\rm out}=R_{\rm H}\equiv a(M_c/3M_\odot)^{1/3}$. Although $R_{\rm rcb}\sim R_{\rm H}$, up to a logarithmic correction, the strong dependence $t_{\rm circ}\propto R_{\rm rcb}^{-5}$ necessitates a more accurate estimate. Using Equation \eqref{eq:radius_rho_d} we find that
\begin{equation}
\frac{R_{\rm H}}{R_{\rm rcb}}=1+\frac{R_{\rm H}}{R_{\rm B}}\ln\frac{\rho_{\rm rcb}}{\rho_d}\approx 2,
\end{equation}
with the numerical value estimated for $5-10M_\oplus$ planets with atmospheres of a few percent in mass at $a=0.1\textrm{ AU}$ \citepalias[see][and substitute $R_{\rm B}\equiv GM_c\mu/k_{\rm B}T_d$ for an accurate solution]{Ginzburg2016}.
We proceed with the calculation of $t_{\rm circ}$ by substituting $R_{\rm rcb}\approx 0.5R_{\rm H}$ in Equation \eqref{eq:t_circ_atmosphere}
\begin{equation}\label{eq:t_circ_super}
t_{\rm circ}\approx 2\frac{Q}{f}\left(\frac{M_\odot}{M_c}\right)^{2/3}P.
\end{equation}

\begin{figure}
	\includegraphics[width=\columnwidth]{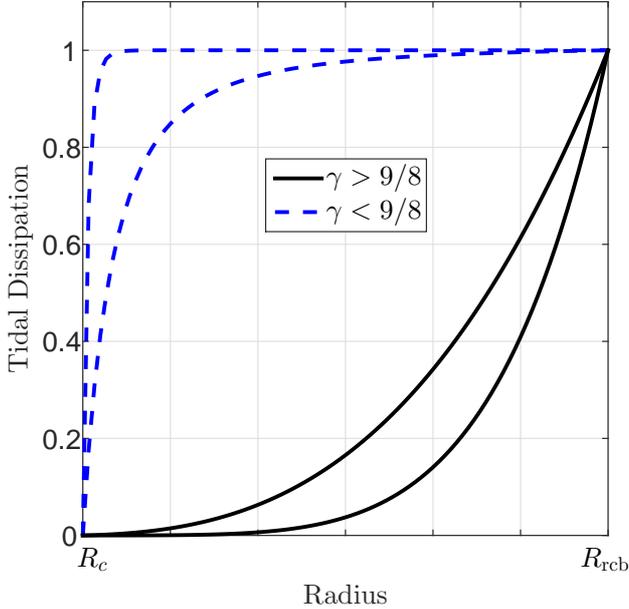}
	\caption{Cumulative tidal dissipation profile for different values of the adiabatic index $\gamma$ (top to bottom: 1.05, 1.1, 1.2, 1.4). The power dissipation below a radius $r$ is given as a fraction of the total power dissipated in the atmosphere (the tidal dissipation in the rocky core is excluded; see text for details). For $\gamma>9/8$ the dissipation is dominated by $r\approx R_{\rm rcb}$ ($=10R_c$ in the figure), while for $\gamma<9/8$ it is concentrated at $r\approx R_c$.}
	\label{fig:profile}
\end{figure}

For an effective $\gamma<4/3$ we integrate Equation \eqref{eq:l_atm} over the entire atmosphere $L_{\rm circ}\propto\int{r^5dm}=4\pi\int{\rho r^7dr}$ with the density profile given by Equation \eqref{eq:rho_convective}. We find that although the mass is now concentrated in the inner atmosphere ($r\sim R_c$), the tidal dissipation is still dominated by the outer layers ($r\sim R_{\rm rcb}$) as long as $\gamma>9/8$, as demonstrated in Fig. \ref{fig:profile}. The values of $\gamma$ found by \citet{Lee2014} and \citet{Piso2015} satisfy this condition. In this case, the atmosphere mass fraction $f$ in the equations above should be multiplied by $(R_c/R_{\rm rcb})^{(4-3\gamma)/(\gamma-1)}$, which comes from the difference between Equations \eqref{eq:mass_atm_outside} and \eqref{eq:mass_atm_inside}. Intuitively, this is the fraction of the atmosphere's mass that extends to large radii, where it contributes significantly to the tidal dissipation. This fraction depends on the deviation of $\gamma$ from 4/3. \citet{Lee2014} find a minimum of $\gamma=1.22$, with most of the convective region having values of $\gamma$ closer to 4/3 (see their Figure 3, with $\nabla_{\rm ad}\equiv 1-1/\gamma$). For the typical values of $\gamma$ found by \citet{Lee2014}, this correction term amounts to a factor of a few, which we disregard here for simplicity.

\section{Effect of Heating on Accretion}\label{sec:effect}

As the planet accretes gas via Kelvin--Helmholtz cooling, $M_{\rm atm}$ increases, and with it the tidal heating $L_{\rm circ}\propto f\equiv M_{\rm atm}/M_c$, according to Equation \eqref{eq:l_atm}. The heat that the planet can radiate away, on the other hand, decreases as the atmosphere becomes more optically thick $L\propto\rho_{\rm rcb}^{-1}\propto M_{\rm atm}^{-1}$, according to Equation \eqref{eq:lum}, and neglecting the mild increase of the opacity with density \citep{Freedman2008,Freedman2014}. Consequently, $M_{\rm atm}$ will reach a critical value, beyond which the atmosphere can no longer cool. We assume here that the tides dissipate their energy in the convective interior of the atmosphere \citep{Liu2008,IbguiBurrows2009} and that the decay of the dissipation with depth is not too steep. Otherwise, if the heat deposition is shallow enough, the atmosphere can continue to cool and grow (but at a slower rate) beyond the critical $M_{\rm atm}$ by developing an alternating radiative--convective structure \citep[see][for details]{GinzburgSari2015,GinzburgSari2016}. This assumption is studied in detail in Appendix \ref{sec:profile}.

We now derive the conditions for the tides to interfere with the atmosphere accretion. The first condition is $t_{\rm circ}>t_{\rm disk}$. Otherwise, the tides prevent atmosphere accretion only for a short time (in which $L$ remains high to evacuate the heat), after which the core has still $\sim t_{\rm disk}$ to gain an atmosphere. The second condition is that the maximum tidal dissipation (achieved at $t=t_{\rm disk}$, when $f$ is maximal), $L_{\rm circ}=E_{\rm circ}/t_{\rm circ}$, is larger than the minimum luminosity (also at $t=t_{\rm disk}$), $L=E_{\rm atm}/t_{\rm disk}$. In the marginal case of $L(t_{\rm disk})=L_{\rm circ}(t_{\rm disk})$, the tides will stop accretion at $t=t_{\rm disk}$. If $L_{\rm circ}>L$, the accretion will stop earlier, with smaller final atmosphere masses (an example is given below). The combined condition for tides to have an effect is
\begin{equation}\label{eq:condition}
1<\frac{t_{\rm circ}}{t_{\rm disk}}<\frac{E_{\rm circ}}{E_{\rm atm}}.
\end{equation}
Intuitively, if the tides are too strong (short $t_{\rm circ}$) they circularize the orbit too early and stop, but if they are too weak (long $t_{\rm circ}$), their dissipated power is too low. We follow \citet{LeeChiang2015} and take $E_{\rm atm}=GM_c M_{\rm atm}/R_c$, suitable for $\gamma<4/3$, for which both the energy and the mass of the atmosphere are concentrated in the inner layers \citepalias[see also][]{Ginzburg2016}. 

In Appendix \ref{sec:appendix} we show that tidal synchronization cannot affect the planet's cooling because $E_{\rm syn}\lesssim E_{\rm atm}$. The energy released by circularization, on the other hand, has nothing to do with the atmosphere's binding energy. From Equation \eqref{eq:e_circ} we see that $E_{\rm circ}>E_{\rm atm}$ if
\begin{equation}\label{eq:condition_energy}
e^2>\frac{M_{\rm atm}}{M_\odot}\frac{a}{R_c}.
\end{equation}
For a nominal $M_c=5M_\oplus$ planet with $f=5\%$ in an $a\approx 0.1\textrm{ AU}$ (equivalently $\approx 10\textrm{ day}$) orbit, Equation \eqref{eq:condition_energy} is satisfied for $e>0.03$. Now we need to check whether the circularization timescale complies with the condition of Equation \eqref{eq:condition}. For our nominal planet, Equation \eqref{eq:t_circ_super} predicts $t_{\rm circ}\approx 10^8\textrm{ yr}$, so we may safely assume that $t_{\rm circ}\gg t_{\rm disk}$ (for lighter envelopes, the time becomes even longer). The fact that $t_{\rm circ}<\textrm{Gyr}$ means that planets can resume their cooling after $t_{\rm circ}$ and have enough time to contract to their present-day observed sizes \citep[see, e.g.,][]{HoweBurrows2015,InamdarSchlichting2016}. In this post-nebular phase, cooling is equivalent to contraction, rather than accretion (since the gas disk has dispersed),  initially with the same luminosity \citepalias{Ginzburg2016}. Finally, using Equations \eqref{eq:e_circ}, \eqref{eq:t_circ_super}, and \eqref{eq:condition} we derive the condition that determines if tides are strong enough to stop atmosphere accretion
\begin{equation}\label{eq:condition_period}
e^2>Q\frac{P}{t_{\rm disk}}\left(\frac{M_c}{M_\odot}\right)^{1/3}\left(\frac{a}{R_c}\right)=Q\frac{P}{t_{\rm disk}}\left(G\rho_c P^2\right)^{1/3},
\end{equation}
with $\rho_c$ denoting the density of the rocky core. Rocky cores are slightly compressed by their high pressure $\rho_c\propto M_c^{1/4}$ \citep[e.g.][]{Valencia2006,Lopez2012}. However, due to the weak relation $e\propto\rho_c^{1/6}$ that we find in Equation \eqref{eq:condition_period}, we simply take a constant $\rho_c\approx 5\textrm{ g cm}^{-3}$. We rewrite Equation \eqref{eq:condition_period} as
\begin{equation}\label{eq:condition_period_final}
e>\left(Q\frac{P}{t_{\rm disk}}\right)^{1/2}\left(\frac{P}{t_{\rm dyn}}\right)^{1/3}\sim 0.2\left(\frac{P}{10\textrm{ day}}\right)^{5/6},
\end{equation}
with the core's dynamical time given by $t_{\rm dyn}\equiv(G\rho_c)^{-1/2}\approx 0.5\textrm{ h}$, and with $t_{\rm disk}=3\textrm{ Myr}$. For a given eccentricity, $P_{\rm crit}/10\textrm{ day}=(e/0.2)^{6/5}$ marks the critical period below which tides influence gas accretion. We can calculate the final atmosphere mass for $P<P_{\rm crit}$ by remembering that $L\propto M_{\rm atm}^{-1}$ and $L_{\rm circ}\propto M_{\rm atm}$. Using these relations and Equation \eqref{eq:condition_period} we find
\begin{equation}\label{eq:m_atm_below}
\frac{M_{\rm atm}}{M_{{\rm atm},0}}=\begin{cases}
\left(P/P_{\rm crit}\right)^{5/6} & P<P_{\rm crit} \\
\quad 1 & P>P_{\rm crit}
\end{cases},
\end{equation}
with $M_{{\rm atm},0}$ marking the mass that the atmosphere would have reached during the disk's lifetime, in the absence of tides. 

Note that, in general, $M_{{\rm atm},0}$ is a function of the core mass, the ambient temperature, and the opacity, and it is calculated by a cooling model, as briefly presented in Section \ref{sec:structure}. For completeness, we provide the following estimate from \citetalias{Ginzburg2016} for $t_{\rm disk}=3\textrm{ Myr}$
\begin{equation}\label{eq:m_atm0}
f_0\equiv\frac{M_{{\rm atm},0}}{M_c}\approx 0.12\left(\frac{M_c}{5M_\oplus}\right)^{0.8}\left(\frac{P}{10\textrm{ day}}\right)^{1/14}.
\end{equation}
In the derivation of Equation \eqref{eq:m_atm0} we relate the disk's temperature to the semi-major axis with $T_d/10^3\textrm{ K}=(a/0.1\textrm{ AU})^{-3/7}$ \citep{ChiangGoldreich1997,Lee2014} and the period to $a$ by assuming a solar-mass star.
By combining Equations \eqref{eq:condition_period_final}, \eqref{eq:m_atm_below}, and \eqref{eq:m_atm0} we find the maximum gas mass fraction a rocky core may accrete when tides are taken into account so that $f=\min (f_0,f_{\rm max})$ with
\begin{equation}\label{eq:f_max}
f_{\rm max}\approx\frac{2\%}{e}\left(\frac{M_c}{5M_\oplus}\right)^{0.8}\left(\frac{P}{10\textrm{ day}}\right)^{19/21}.
\end{equation}

To summarize, planets in orbits longer than $P_{\rm crit}(e)$ are not affected by tidal dissipation, and their gas mass is determined by the accretion time ($t_{\rm disk}$). The cooling and accretion in planets with shorter orbits stop before the disk disperses, preventing their atmospheres from reaching their full mass potential, as seen in Equation \eqref{eq:m_atm_below}. The final mass of these atmospheres if given by Equation \eqref{eq:f_max}.

\subsection{Eccentricity Origins}\label{sec:ecc}

We demonstrated above that the accretion from the nebula may be affected by tidal heating if planets have high eccentricities during this phase. While a detailed study of the origins of such eccentricities is beyond the scope of this work, we notice that the timescale for eccentricity evolution due to planet--disk interaction is very short. From \citet{GoldreichTremaine1980} and \citet{GoldreichSari2003} this timescale is given by
\begin{equation}\label{eq:t_ecc}
t_e=\left(e\frac{a}{H}\right)^3\left(\frac{\Delta a}{a}\right)^4\frac{M_\odot}{M_c}\frac{M_\odot}{M_{\rm disk}} P\sim e^3\left(\frac{a}{R_{\rm B}}\right)^{1/2}\left(\frac{M_\odot}{M_c}\right)^{3/2}P,
\end{equation}
with $M_{\rm disk}$ marking the gas disk's mass enclosed inside a radius $a$. We estimate $M_{\rm disk}\sim M_c$ by assuming that massive cores form only once the gas density has decreased to the order of the solid density, which forms the core \citep[only then can protocores collide, e.g.,][]{Goldreich2004b}. We remark that for widely-separated protocores the critical disk mass is $M_{\rm disk}\ll M_c$, significantly increasing $t_e$ \citep[see][and references within]{Dawson2016,LeeChiang2016}.
The minimum distance of the interacting gas $\Delta a$ is taken as the gap width for gap-opening planets \citep{GoldreichSari2003}. Our nominal $5M_\oplus$ planet marginally opens a gap \citep{Rafikov2002,DuffellMacfadyen2013}, so we substitute $\Delta a\sim H$ in Equation \eqref{eq:t_ecc}, with $H\sim (k_{\rm B}T_d/\mu)^{1/2}P$ denoting the disk's vertical scale height. The $(ea/H)^3$ factor in Equation \eqref{eq:t_ecc} is adequate for $e>H/a$, which is the case for the high eccentricities required in our scenario \citep{PapaloizouLarwood2000}. For our nominal $M_c=5M_\oplus$ and $P=10\textrm{ days}$ (equivalently $a\approx 0.1\textrm{ AU}$) planet, Equation \eqref{eq:t_ecc} yields $t_e\sim 10^4\textrm{ yr}\ll t_{\rm disk}$ (though see the remark concerning $M_{\rm disk}$ above).

The short $t_e$ that we obtain implies that eccentricities must be continuously excited during the nebula's lifetime, or else they are efficiently damped by the gas. Interestingly, several studies find that planet--disk interactions may excite, rather than damp, orbital eccentricity with a similar timescale $t_e$ \citep{GoldreichSari2003,DuffellChiang2015,TeyssandierOgilvie2016}. Another possibility is planet--planet interactions (many of the observed low-density planets are found in packed multi-planet systems). We note that if the eccentricity is continuously pumped up during the disk's lifetime, the criterion $t_{\rm circ}>t_{\rm disk}$ (see discussion above) is redundant.

\section{Relation to Observations}\label{sec:observations}

\begin{figure}
	\includegraphics[width=\columnwidth]{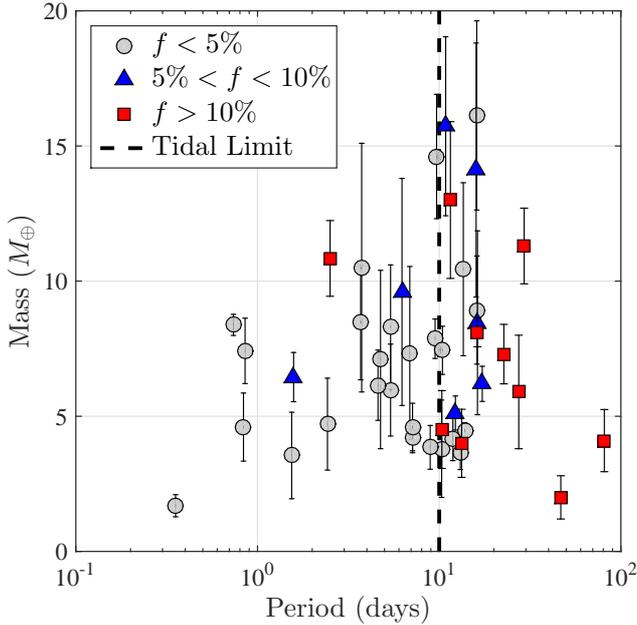}
	\caption{Observed super-Earth population \citep[][see text for details]{WeissMarcy2014,JontofHutter2016}. The planets are grouped according to their gas mass fraction $f$, estimated by the same method as in \citetalias{Ginzburg2016}. Low-density planets are marked with blue triangles ($5\%<f<10\%$) or red squares ($f>10\%$). 
	The tidal limit (dashed black line) is according to Equation \eqref{eq:condition_period_final} with an eccentricity $e=0.2$. Below it we expect low-density planets to be less common, according to Equation \eqref{eq:m_atm_below}.}
	\label{fig:obs}
\end{figure}

In Fig. \ref{fig:obs} we present a scatter plot of transiting short-period super-Earths ($R<4R_\oplus$) with masses known to within 50\% (measured from radial velocity or transit time variations) from \citet{WeissMarcy2014} and  \citet{JontofHutter2016}. Each planet's gas mass fraction $f$ is estimated from its mass and radius by the same method used in \citetalias{Ginzburg2016}, which yields similar results to \citet{Lopez2012}. It is evident from the figure that low-density planets (coloured squares and triangles) are found almost exclusively beyond $\approx 10$ day orbits (equivalently $\approx 0.1\textrm{ AU}$). These are planets that require gas atmospheres to explain their large radii \citep[$f\gtrsim 5\%$; see, e.g.,][]{Lopez2012}. This is in agreement with \citet{Youdin2011} who finds a deficit of $\sim 3R_\oplus$ planets (the mass is usually unknown) in orbits shorter than 7 days. These are most likely low-density planets with $f$ of a few percent \citep{LopezFortney2014,WeissMarcy2014,Rogers2015,WolfgangLopez2015}. 

Atmospheric mass-loss models, usually due to high-energy stellar photons, have been successful in explaining this paucity of low-density super-Earths in close proximity to the star, where the UV flux is high \citep{Lopez2012,OwenWu2013,Ginzburg2016,Lundkvist2016}.
Nonetheless, Fig. \ref{fig:obs} demonstrates that tidal heating may play an important role in shaping the observed planet population as well. Specifically, the 10 day threshold is naturally reproduced if the planets had initial eccentricities of the order of 0.2. Moreover, the limit set by tides can prevent the runaway growth of super-Earths into gas giants if accretion is as efficient as suggested by \citet{Lee2014}. Such  eccentricities are much higher than the ones we observe today \citep[e.g.][]{Lissauer2013,WuLithwick2013}, consistent with the short circularization timescale ($t_{\rm circ}\sim 10^8\textrm{ yr}$) found in Section \ref{sec:effect}.
 
A detailed study of possible planet migration is beyond the scope of this work, which has assumed that atmosphere accretion occurred in situ (the core could have migrated). It is clear that inward migration of fully formed planets (after the accretion and tidal heating have ceased) can bring low-density planets from long to short periods, where their in situ formation is impossible due to the strong tides. The outcome of possible dynamical instability of multi-planet systems with high eccentricities is also not considered here. 

\section{Conclusions}\label{sec:conclusions}

Many observed super-Earths probably have voluminous gas atmospheres that constitute a few percent of their mass. Explaining these atmospheres by gas accretion from the surrounding protoplanetary disk is not trivial. If the accretion rate is too low, rocky cores do not accrete enough gas before the disk disperses. On the other hand, if the accretion is too fast, the planets evolve into Jupiters. While some studies find that the accretion rates and typical disk lifetimes are consistent with such atmospheres \citepalias{Ginzburg2016}, other studies \citep{Lee2014,LeeChiang2016} find that accretion is somewhat too efficient and suggest different mechanisms that delay (cores form by late collisions of smaller protocores) or slow it down \citep[higher opacities or atmosphere replenishment, see, .e.g.,][]{Ormel3D}.  

In this work we studied an alternative mechanism that may prevent the runaway growth of accreting super-Earths. Since the bottleneck for gas accretion is the rate by which the infalling gas can radiate away its gravitational energy, tidal heating of the atmosphere can inhibit its cooling and further accretion. Conceptually, this mechanism is similar to the suppression of hot-Jupiter cooling by tidal dissipation and of super-Earth atmosphere accretion by heating from  planetesimal impacts, discussed in previous studies (see Section \ref{sec:introduction}).

During the nebular accretion phase, which lasts a few Myr, super-Earth atmospheres extend to their Hill radius $R_{\rm H}$, making them larger than Jupiter. We find that the tidal dissipation $L_{\rm circ}$ in such young planets is dominated by the inflated atmosphere and grows as the gas mass accumulates $L_{\rm circ}\propto M_{\rm atm}$. The heat that the planet can radiate (i.e. the internal luminosity), on the other hand, decreases as the atmosphere becomes more massive and optically thick $L\propto M_{\rm atm}^{-1}$. Thus, the atmosphere will reach a maximal mass which satisfies $L=L_{\rm circ}$. 

Quantitatively, we find that super-Earth atmospheres reach a mass $M_{\rm atm}/M_{{\rm atm},0}=(P/P_{\rm crit})^{5/6}$, where $M_{{\rm atm},0}$ is the mass a planet accretes from the disk during its lifetime $t_{\rm disk}$ in the absence of tidal heating, $P$ is its orbital period, and $P_{\rm crit}$ is the critical period beyond which planets are not affected by tidal heating ($M_{\rm atm}=M_{{\rm atm},0}$ for $P>P_{\rm crit}$). The tidal limit is given by $(QP_{\rm crit}/t_{\rm disk})^{1/2}(P_{\rm crit}/t_{\rm dyn})^{1/3}=e$, where $Q$ is the tidal dissipation parameter, $t_{\rm dyn}$ is the dynamical timescale of the rocky core, and $e$ the initial eccentricity. For $e\sim 0.2$, $P_{\rm crit}\approx 10\textrm{ days}$, potentially explaining why super-Earths with such periods did not become gas giants. Furthermore, this tidal limit can naturally account for the dearth of low-density super-Earths on shorter orbits. A possible origin of such high eccentricities is planet--disk interaction (see Section \ref{sec:ecc}). 

While other mechanisms, such as atmosphere evaporation by stellar UV radiation \citep{Lopez2012,OwenWu2013,Lundkvist2016}, are usually invoked in order to explain the observed super-Earth distribution, we demonstrated here that tidal dissipation may play an equally important role. The main difference between the two mechanisms, which may allow us to differentiate between them, is that UV evaporation sculpts the planet population after the nebula vanishes, by blowing away existing envelopes, whereas tidal heating limits the accretion of gas atmospheres in the first place, while the nebula is still present. Another difference is the effect of the planet's mass. UV evaporation is significantly less efficient for massive planets due to their deep potential wells \citep{LopezFortney2013,Ginzburg2016}. We find that tidal heating is less sensitive to the mass (though $M_{{\rm atm},0}$ itself is a function of the mass). Admittedly, at least in some cases, this mass dependence seems to support the evaporation scenario \citep{LopezFortney2013}. An additional method to differentiate between the two models is to distinguish orbital period from UV flux using accurate host-star characterization \citep{Lundkvist2016}. 

In summary, we have demonstrated that tidal heating may play an important role in shaping the observed super-Earth population by setting a separation-dependent limit on the atmosphere mass that they can attract from the nebula. Future studies and observations can break the degeneracy between this and other mechanisms and shed light on the formation of these intriguing planets.

\section*{Acknowledgements}

This research was partially supported by ISF, ISA and iCore grants. We thank the Harvard--Smithsonian Center for Astrophysics and the Radcliffe Institute for Advanced Study for warm hospitality during the initial stages of the research. We thank Hilke Schlichting, Andrew Youdin, and especially the anonymous referee for valuable comments that improved the paper.




\bibliographystyle{mnras}
\bibliography{tides} 



\appendix

\section{Tidal Synchronization}\label{sec:appendix}

Synchronizing tides can be studied as damped harmonic oscillators forced by the relative rotation $\omega$ of the planet with respect to its orbit around the star. The response of such an oscillator (i.e. the raised tides) lags behind the forcing with a phase difference $\phi\approx\tan\phi\sim Q^{-1}$.
The torque exerted by the tidal force on the tidal bulge is
\begin{equation}\label{eq:torque}
N=\frac{GM_\odot M_{\rm atm}}{a^2}\frac{R_{\rm rcb}}{a}z\sin\phi\approx Q^{-1}f\frac{GM_\odot^2}{R_{\rm rcb}}\left(\frac{R_{\rm rcb}}{a}\right)^6,
\end{equation}
where the tides are dominated by the atmosphere due to the $N\propto mr^5/Q$ dependence, as in Section \ref{sec:heating}.

The power dissipated by synchronizing torques is given by $L_{\rm syn}=\omega N$ and the synchronization timescale is
\begin{equation}\label{eq:t_synch}
t_{\rm syn}=\frac{I\omega}{N}=Q\omega\frac{R_{\rm rcb}^3}{GM_c}\left(\frac{M_c}{M_\odot}\right)^2\left(\frac{a}{R_{\rm rcb}}\right)^6,
\end{equation} 
with $I$ denoting the planet's moment of inertia. For $f\gtrsim 1\%$ discussed in this work, $I\propto mr^2$ is also dominated by the atmosphere.
Equation \eqref{eq:t_synch} shows that hot Jupiters (for which simply $R_{\rm rcb}=R$ and $M_c=M$ due to their roughly uniform density) with orbits of $a\approx 0.05\textrm{ AU}$ become tidally locked after a few Myr, assuming $\omega$ is initially equal to Jupiter's current rotation rate and substituting $Q\sim 10^5$ \citep[see, e.g.,][]{Guillot1996}. Therefore, heat due to synchronizing tides cannot explain the puzzling inflation of many observed hot Jupiters. Only a heat source that persists for Gyrs can keep these planets inflated today by hindering their cooling. For this reason, many studies invoke circularizing tides, due to a non-zero eccentricity which is possibly excited by perturbing planets \citep{Bodenheimer2001,Bodenheimer2003,Gu2003}, or obliquity tides \citep{WinnHolman2005}. As we explain below, a similar problem arises in the context of super-Earth atmosphere accretion, leading us to discuss circularizing tides as well.

Quantitatively, as explained in Section \ref{sec:heating}, the synchronization timescale of young super Earths is given by substituting $R_{\rm rcb}\sim R_{\rm H}$ in Equation \eqref{eq:t_synch}. The final step in the calculation of $t_{\rm syn}$ is to estimate $\omega$. The natural scale for the spin is the break-up velocity $\omega^2\sim GM_c/R_{\rm H}^3\sim\omega_{\rm kep}^2$, with $\omega_{\rm kep}\sim P^{-1}$. Obviously, the spin cannot exceed this value. In addition, this is the relative angular velocity of infalling gas at the outer boundary due to the Keplerian shear ($\omega_{\rm kep}^2=GM_\odot/a^3$). We substitute $R=R_{\rm H}$ and $\omega=\omega_{\rm kep}$ in Equation \eqref{eq:t_synch} and find that $t_{\rm syn}\sim QP\ll t_{\rm disk}$ where we assume a nominal $P\approx 10\textrm{ days}$, a disk lifetime $t_{\rm disk}$ of a few Myr (see Section \ref{sec:introduction}) and $Q\sim 10^5$ (see Section \ref{sec:heating}).

We immediately notice that due to Equation \eqref{eq:condition}, and since $t_{\rm syn}\ll t_{\rm disk}$, synchronization does not affect atmosphere accretion. In fact, this statement is not merely a dissipation timescale issue but a more robust energy argument. From Equation \eqref{eq:condition} we require $E_{\rm syn}>E_{\rm atm}$, regardless of the tidal timescale. The spin of the planet simply does not store more energy than the gravitation (assuming break-up rotation) $E_{\rm syn}=I\omega^2\lesssim E_{\rm atm}$, so the cooling planet has no difficulty to evacuate any heat due to synchronization.

\section{Tidal Dissipation Profile}\label{sec:profile}

In Section \ref{sec:effect} we found the maximal atmosphere mass fraction $f_{\rm max}$ that a rocky core can acquire, restricted by the tidal dissipation heat $L_{\rm circ}$ that has to be evicted from the planet. However, as emphasized in the theory of hot-Jupiter inflation by heat deposition \citep[see][and references within]{SpiegelBurrows2013,GinzburgSari2015}, the dissipation depth plays a crucial role in determining the effects of the dissipated power on the planet's evolution. Specifically, we assumed that once the internal luminosity satisfies $L<L_{\rm circ}$, cooling and mass accretion stop. As explained in \citet{GinzburgSari2016}, this assumption holds only if the power deposition profile decays moderately enough with depth so that
it triggers convection (in order to evict the heat) reaching to the bottom of the atmosphere. In this section we study the deposition profile with depth and test this assumption.

We start by calculating the radiative profile required to radiate the tidal dissipation heat, and then check if this profile is stable against convection. By combining the hydrostatic equilibrium equation ${\rm d}p/{\rm d}r\propto -\rho/r^2$ ($p$ is the pressure) with the diffusion equation ${\rm d}T/{\rm d}r\propto -\kappa\rho L/(T^3 r^2)$ we find that
\begin{equation}\label{eq:dt_dp}
\frac{{\rm d}T}{{\rm d}p}\propto\frac{\kappa L}{T^3}.
\end{equation}
We make an ansatz that the radiative profile is of the form $p\propto\rho^\Gamma$, with some power $\Gamma$. With this assumption, and utilizing the results of Sections \ref{sec:structure} and \ref{sec:heating} (with $\Gamma$ instead of $\gamma$), we write the tidal dissipation power as $L\propto mr^5\propto\rho r^8\propto r^{(8\Gamma-9)/(\Gamma-1)}\propto T^{-(8\Gamma-9)/(\Gamma-1)}$. For a constant opacity $\kappa$, Equation \eqref{eq:dt_dp} yields $p\propto T^{(12\Gamma-13)/(\Gamma-1)}$. On the other hand, from the ideal gas law $p\propto T^{\Gamma/(\Gamma-1)}$, resulting in $\Gamma=13/11\approx 1.18$. 

For $\gamma\approx 1.2>\Gamma$ \citep{LeeChiang2015}, the radiative profile above is (marginally) stable against convection, since an adiabatic temperature profile is steeper. However, as seen in Equation \eqref{eq:dt_dp}, a slight increase of the opacity with depth \citep[temperature or pressure, see, e.g.,][]{Freedman2008,Freedman2014} triggers convection, justifying the assumptions of this work.

\begin{figure}
	\includegraphics[width=\columnwidth]{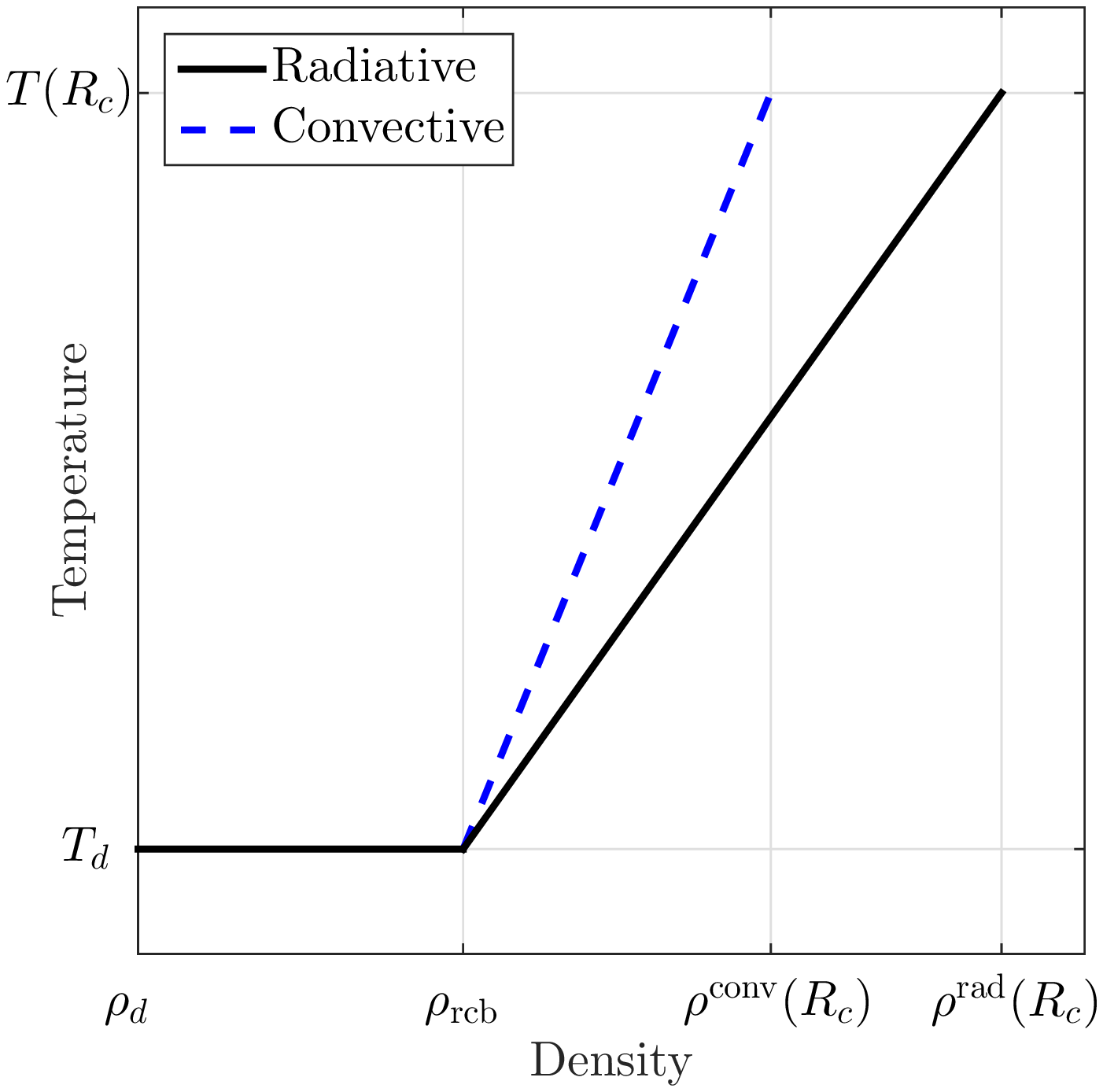}
	\caption{Schematic radiative equilibrium profile (solid black line) induced by tidal dissipation, for the $\Gamma<\gamma$ case, for which the radiative profile is stable against convection (see text for details). A convective profile (dashed blue line) is also plotted for comparison.}
	\label{fig:scheme_rad}
\end{figure}

Nonetheless, we also provide, for completeness, a correction to $f_{\rm max}$ in the case of a radiative profile. In Figure \ref{fig:scheme_rad} we schematically present the difference between convective and radiative profiles for the $\Gamma<\gamma$ case. As seen in the figure, in this case the radiative atmosphere can contract to a larger density in comparison with a convective profile (the temperature profile is roughly the same, as explained in Section \ref{sec:structure}). Quantitatively, the atmosphere's maximal mass, which is dominated by the density at the bottom of the atmosphere $\rho(R_c)$ for $\gamma<4/3$ (see Section \ref{sec:structure}), increases by a factor of
\begin{equation}
\frac{f_{\rm max}^{\rm rad}}{f_{\rm max}^{\rm conv}}=\frac{\rho^{\rm rad}(R_c)}{\rho^{\rm conv}(R_c)}\sim\left(\frac{R_{\rm B}'}{R_c}\right)^{1/(\Gamma-1)-1/(\gamma-1)},
\end{equation}
where we use the relation $T(R_c)/T_d\sim R_{\rm B}'/R_c$ according to Equations \eqref{eq:bondi} and \eqref{eq:tmp_core}.
For a constant $\kappa$, this is a factor of a few, since $\Gamma$ and $\gamma$ are similar, as derived above. We stress again that for realistic opacities, $\Gamma>\gamma$, and there is no correction factor because the atmosphere's profile is convective.

\bsp	
\label{lastpage}
\end{document}